\documentclass[11pt]{article}\usepackage{epsfig}
\begin{document}
\begin{titlepage}

\hfill {SNUTP/00-014}

\begin{center}
\ \\
{\Large \bf The Gluon Spin in the Chiral Bag Model}
\ \\
\vspace{.3cm}
{Hee-Jung Lee and Dong-Pil Min }

{\it Department of Physics and Center for Theoretical Physics}\\
{\it Seoul National University, Seoul 151-742, Korea}\\ ({\small
E-mail: hjlee@fire.snu.ac.kr; dpmin@phya.snu.ac.kr})

\vskip 0.2cm
{Byung-Yoon Park}

{\it Department of Physics, Chungnam National University, Daejon 305-764,
Korea}\\ ({\small E-mail: bypark@chaosphys.chungnam.ac.kr})

\vskip 0.2cm
{Mannque Rho}

{\it Service de Physique Th\'eorique, CE Saclay}\\ {\it 91191
Gif-sur-Yvette, France}\\ {\&}\\{\it School of Physics, Korea
Institute for Advanced Study, Seoul 130-012, Korea}\\ ({\small
E-mail: rho@spht.saclay.cea.fr})


\vskip 0.2cm
{Vicente Vento }

{\it Departament de Fisica Te\`orica and Institut de F\'{\i}sica
Corpuscular}\\ {\it Universitat de Val\`encia and Consejo Superior
de Investigaciones Cient\'{\i}ficas}\\ {\it E-46100 Burjassot
(Val\`encia), Spain}\\ ({\small E-mail: vicente.vento@uv.es})

\end{center}
\vskip 0.2cm \centerline{\bf Abstract} We study the gluon
polarization contribution at the quark model renormalization scale
to the proton spin,  $\Gamma$, in the chiral bag model. It is
evaluated by taking the expectation value of the forward matrix
element of a local gluon operator in the axial gauge $A^+=0$. It
is shown that the confining boundary condition for the color
electric field plays an important role. When a solution satisfying
the boundary condition for the color electric field, which is not
the conventionally used but which we favor, is used,
 the $\Gamma$ has a positive
value for {\it all} bag radii and its magnitude is comparable to
the quark spin polarization. This results in a significant
reduction in the relative fraction of the proton spin carried by
the quark spin, which is consistent with the small flavor singlet
axial current measured in the EMC experiments. \vskip 0.25cm

\leftline{Pacs: 12.39Ba, 12.39Dc,14.20Dh,14.70Dj} 

\leftline{Keywords: quark, gluon, polarization, 
chirality, bag model}

\end{titlepage}
The EMC experiment\cite{EMC} revealed the surprising fact that
less than 30\% of the proton spin may be carried by the quark
spin. This is at variance with what one expects from 
non-relativistic or relativistic constituent 
quark models. This discrepancy -- so called ``the proton spin
crisis" -- can be understood as an effect associated with the axial
anomaly\cite{TJZ}. For example it has been argued \cite{AEL} 
that the  experimentally measured quantity is not merely the quark spin
polarization $\Delta\Sigma$ but rather the flavor singlet axial
charge, to which the gluons contribute through the axial anomaly.
Another interpretation \cite{CH96} is that the anomalous gluons can
induce a sea-quark polarization through the axial anomaly, which
cancels the spin from the valence quarks,
if the gluon spin component is positive. These explanations, and
possibly others, could be reconciled if one would establish that they
are gauge dependent statements, while the
measured quantity is gauge-invariant~\cite{cheng2}.

Although not directly observable, an equally
interesting quantity related to the proton spin is the fraction of
spin in the proton that is carried by the gluons. In
Ref.\cite{Jaf96}, the gluon spin $\Gamma$ is introduced as
\begin{equation}
\textstyle \frac12 = \frac12 \Sigma +  L_Q + \Gamma + L_G,
\label{QG}
\end{equation}
where $L_{Q,G}$ is the orbital angular momentum of the
corresponding constituent and $\Gamma$ is defined as the integral
of the polarized gluon distribution in analogy to $\Sigma$. The spin
of course is gauge-invariant but the individual components in
(\ref{QG}) may not be. $\Gamma$ can be
expressed as a matrix element of products of the gluon vector
potentials and field strengths in the nucleon rest frame and in
the $A^+=0$ gauge. When evaluated with the gluon fields
responsible for the $N-\Delta$ mass splitting,  $\Gamma$ turns
out to be negative, $\Gamma\sim -0.1
\alpha_{\mbox{\scriptsize{bag}}}$, in the MIT bag model and even
more so in the non-relativistic quark model.

By contrast, there are several other calculations that give
results with  opposite sign. For example, the QCD sum rule
calculation\cite{MPS97} yields a positive value $2\Gamma \sim 2.1
\pm 1.0$ at 1 $\mbox{GeV}^2$. In Ref.\cite{BCD98}, it is suggested
that the negative $\Gamma$ of Ref.\cite{Jaf96} could be due to
neglecting ``self-angular momentum."  The authors of \cite{BCD98}
show that when self-interaction contributions are included, one
obtains a positive value $\Gamma\sim +0.12$ in the Isgur-Karl
quark model at the scale $\mu^2_0 \approx 0.25\mbox{ GeV}^2$. Once
the gluons contribute a significant fraction to the proton spin,
due to the normalization Eq.(1), the relative
fraction of the proton spin lodged in the quark spin changes. Thus, the
positive gluon spin seems to be consistent with the EMC
experiment.

In this letter, we address this issue in the chiral bag model
and pay special attention to the confining boundary condition
for the gluon fields.

Let us start by briefly reviewing how the gluon spin operator is
derived in Ref.\cite{Jaf96,JM90}. From the Lagrangian
\begin{equation}
{\cal L}=-\frac{1}{2}{\rm Tr}\bigg(F^{\mu\nu}F_{\mu\nu}\bigg)
\end{equation}
with $F^{\mu\nu}=\frac{\lambda^a}{2}F^{a\mu\nu}$, one gets the
gluon angular momentum tensor
\begin{equation}
M^{\mu\nu\lambda}=2{\rm Tr}\bigg(
x^{\nu}F^{\mu\alpha}{F_{\alpha}}^{\lambda}
-x^{\lambda}{F_{\mu\alpha}}^{\nu}\bigg)
-(x^{\nu}g^{\mu\lambda}-x^{\lambda}g^{\mu\nu}){\cal L}.
\end{equation}
Integrating by parts, we have
\begin{eqnarray}
M^{\mu\nu\lambda}&=&2{\rm Tr}\bigg(
-F^{\mu\alpha}(x^{\nu}\partial^{\lambda}
-x^{\lambda}\partial^{\nu})A_{\alpha}
+F^{\mu\lambda}A^{\nu}+F^{\nu\mu}A^{\lambda} \nonumber\\
&&+\partial_{\alpha}
(x^{\nu}F_{\mu\alpha}A^{\lambda}-x^{\lambda}F_{\mu\alpha}A^{\nu})
+\frac{1}{4}F^{\mu\nu}F_{\mu\nu}
(x^{\nu}g^{\mu\lambda}-x^{\lambda}g^{\mu\nu})\bigg).
\label{M}\end{eqnarray}
It seems reasonable to interpret the
first term as the gluon orbital angular momentum contribution and
the second as that of the gluon spin, while recalling that this is a
gauge dependent statement. We will not consider the fourth
term hereafter, since it contributes only to boosts. In
Ref.\cite{Jaf96,JM90}, the third term is also dropped as is done
in the open space field theory. When finite space is
involved, as in the bag model, dropping this term requires that the
gluon fields satisfy boundary conditions on the surface of the region, 
as we next show. Let us express the gluon  angular momentum operator 
in terms of the Poynting 
vector, i.e.,
\begin{equation}
{\bf J}_G
= 2{\rm Tr}\int_V d^3r[{\bf r}\times({\bf E}\times{\bf B})].
\label{Jg}
\end{equation}
Now doing the partial integration for ${\bf B}=\nabla\times{\bf
A}$, we have
\begin{equation}
J^k_G = 2{\rm Tr}\left\{\int_V d^3r\bigg(E^l({\bf
r}\times\nabla)^k A_l + ({\bf E}\times{\bf A})^k\bigg) -
\int_{\partial V}d^2r({\bf r}\cdot{\bf E})({\bf r}\times{\bf A})^k
\right\}. \label{J^k_g}
\end{equation}
The surface term is essential to make the whole
angular momentum operator gauge-invariant, but the surface
term only vanishes, if the electric field satisfies the boundary
condition on the surface,
\begin{equation}
{\bf r}\cdot{\bf E}=0. \label{bcE}\end{equation}
This is just the MIT boundary condition for gluon confinement.
However, the static electric field traditionally used \cite{MIT75} 
does not satisfy this condition. Instead 
color singlet nature of the hadron states is imposed to assure
confinement globally.

We next show that the negative $\Gamma$ of Ref.\cite{Jaf96} results
if this procedure to confine color is imposed. To proceed, we choose the
$A^+=0$ gauge and write the gluon spin in a local form as
\begin{equation}
\Gamma=\langle p, \uparrow |2{\rm Tr}\int_V d^3x \bigg(
({\bf E}\times{\bf A})^3+{\bf B}_{\perp}\cdot{\bf A}_{\perp}
\bigg) |p, \uparrow\rangle,
\end{equation}
where $\perp$ denotes the direction perpendicular to the proton
spin polarization and the superscript $+$ indicates the light cone
coordinates defined as $x^\pm = \frac{1}{\sqrt 2}(x^0 \pm x^3)$.
We shall evaluate this expression by incorporating the exchange of the
static gluon fields between $i$-th and $j$-th quarks ($i\neq j$)
which are responsible for the $N-\Delta$ mass splitting in the bag
model.

In the chiral bag model, the static gluon fields are generated by
the color charge and current distributions of the
$i$-th valence quark given by\cite{BV90}
\begin{eqnarray}
J^{0a}_i({\bf r}) &=& \frac{g_s}{4\pi} {\rho(r)}
\frac{\lambda^a_i}{2},
\label{J0}\\
{\bf J}^a_i({\bf r}) &=& \frac{g_s}{4\pi}
3 (\hat{\bf r} \times {\bf S}) \frac{\mu^\prime (r)}{r^3}
\frac{\lambda^a_i}{2},
\label{vecJ}
\end{eqnarray}
where $\rho(r)$ and $\mu^\prime (r)$ are, respectively, the quark
number  and current densities determined by the valence quark
wave functions. (See Ref.\cite{BV90} for their explicit formulas.)
They are very similar in form to those of the MIT bag model. There
is, however, an essential difference, namely, that the spin in the
chiral bag model is given by the collective rotation of the whole
system while in the MIT bag it is given by a individual contribution
of each constituent, i.e.,  there is no index $i$ in the spin operator
in Eq.(\ref{vecJ}).

The charge and current densities yield the color electric and
magnetic fields as
\begin{eqnarray}
{\bf E}^a_i &=& \frac{g_s}{4\pi} \frac{Q(r)}{r^2}\frac{\lambda^a_i}{2}
\hat{\bf r},
\label{vecE}\\
{\bf B}^a_i &=& \frac{g_s}{4\pi} \left\{ {\bf S} \left(
2M(r) + \frac{\mu(R)}{R^3} - \frac{\mu(r)}{r^3} \right)
+ 3 \hat{\bf r} (\hat{\bf r}\cdot {\bf S}) \frac{\mu(r)}{r^3}
\right\} \frac{\lambda^a_i}{2},
\label{vecB}
\end{eqnarray}
where
\begin{eqnarray}
\mu(r) &=& \int^r_0 d r^\prime \mu^\prime (r^\prime), \\
M(r) &=& \int^R_r dr^\prime \frac{\mu^\prime (r^\prime)}{r^{\prime 3}}.
\end{eqnarray}
The quantity $Q(r)$ can be determined from Maxwell's equations. The
most general solution can be written as
\begin{equation}
Q_\lambda(r) = 4\pi\int^r_\lambda dr^\prime r^{\prime 2} \rho(r^\prime),
\end{equation}
with an arbitrary $\lambda$. The standard procedure
is to choose $\lambda=0$ so that the electric field is regular at
the origin. This has been the adopted convention in the early days\cite{MIT75}. 
We will refer to this solution as $Q_0(r)$.
However, $Q_0(r)$ does not satisfy the {\it local} boundary
condition (\ref{bcE}), since it is normalized as $Q_0(R)=1$. In
Ref.\cite{MIT75}, the fact that hadrons are color singlet states, had to be
imposed in order to justify the use of this solution.

Another solution is obtained by
setting $\lambda=R$ \cite{BV90} and we will look for its consequences here. 
This choice satisfies the local boundary condition but
requires the relaxation of the continuity of the electric fields inside the
bag. It has been shown in \cite{BV90}, and will be shown here again,
that these two solutions, $Q_0(r)$ and $Q_R(r)$, lead to dramatic
differences for certain observables.

By using the static Green functions and the Coulomb gauge
condition, one can obtain time-independent scalar and
vector potentials from the charge and current densities,
(\ref{J0}) and (\ref{vecJ}),
\begin{eqnarray}
\Phi^a_i({\bf r})&=&\frac{g_s}{4\pi} \frac{Q_\lambda(r)}{r}
\frac{\lambda^a_i}{2},
\label{Phi} \\
{\bf U}^a_i({\bf r})&=&\frac{g_s}{4\pi} h(r)
({\bf S}\times{\bf r}) \frac{\lambda^a_i}{2},
\label{U}
\end{eqnarray}
where
\begin{equation}
h(r) = \bigg( \frac{\mu(R)}{2R^3} + \frac{\mu(r)}{r^3} + M(r) \bigg).
\end{equation}
From these, the appropriate scalar and vector potentials
satisfying the $A^+=0$ gauge condition can be constructed:
\begin{eqnarray}
A^{a0}_i({\bf r}) &=& \Phi^a_i ({\bf r}),
\nonumber\\
{\bf A}^a_i({\bf r}) &=& {\bf U}^a_i({\bf r})
-\nabla \int_0^z d\zeta\  \Phi^a_i(x,y,\zeta),
\end{eqnarray}
where the direction of the proton polarization is taken as
that of the $z$-axis.
Finally, we obtain
\begin{eqnarray}
\Gamma_\lambda
&=& \sum_{i\neq j}\sum_{a=1}^8 \langle p,\uparrow|\bigg(2\int_V d^3x
({\bf E}^a_i \times {\bf U}^a_j )^3
\nonumber\\
&& + \int_{\partial V} d^2s \hat{\bf z} \cdot \hat{\bf r}
\bigg( U_i^{a1}({\bf x}) \int_0^z d \zeta \ E_i^{a2}(x,y,\zeta)
\nonumber\\
&& - U_i^{a2}({\bf x})\int_0^z d\zeta\ E_i^{a1}(x,y,\zeta)\bigg)\bigg)
|p,\uparrow\rangle
\nonumber\\
&=&\frac{4}{3} \alpha \int_0^Rrdr\ Q_\lambda(r) (h(R)-2h(r)),
\end{eqnarray}
where $\alpha=g^2_s/4\pi$.
The numerical factor in front of the final formula
comes from the fact that
$\sum_a \langle \lambda^a_i \lambda^a_j
\rangle_{\mbox{\scriptsize baryon}}= -8/3$
for $i \neq j$ so that
\begin{equation}
\sum_{i\neq j}\sum_{a=1}^8 \langle p, \uparrow |
S_3 \frac{\lambda_i^a}{2} \frac{\lambda_j^a}{2}
|p,\uparrow \rangle = -2,
\end{equation}
and the integration over the angle yields $1/3$.
It is different from 8/9 of the MIT bag model\cite{Jaf96},
which comes from the expectation value
\begin{equation}
\sum_{i\neq j}\sum_{a=1}^8 \langle p, \uparrow |
\sigma_i^3 \frac{\lambda_i^a}{2} \frac{\lambda_j^a}{2}
|p,\uparrow \rangle = - 4/3.
\end{equation}

It is interesting to note that, if we naively substitute the
static gluon fields $\Phi^a_i$ and ${\bf U}^a_i$ of
Eqs.(\ref{Phi}) and (\ref{U}) satifying the Coulomb gauge
condition into the second term of Eq.(\ref{J^k_g}), we get
\begin{equation}
\Gamma^\prime = -\frac{4}{3}\alpha \int_0^R r dr \ Q_\lambda(r)h(r),
\end{equation}
which is the same expresion that was used in Refs.\cite{BV90,HZ89} to
evaluate the anomalous gluon contribution to the flavor singlet
axial current $a_0$ with the extra factor $(-N_f\alpha/2\pi)$,
i.e., $a_0 = \Sigma - (N_f \alpha/2\pi) \Gamma^\prime_\lambda$. On
the other hand, in Ref.\cite{BCD98}, the gluon spin $\Gamma$
instead of $\Gamma^\prime$ is used for the anomaly correction term
because the calculation is performed in the $A^+=0$ gauge.

If the gluons can contribute to the proton spin, then the
collective coordinate quantization scheme of the chiral bag model
has to be modified to incorporate their contribution. That is because
there is a natural sum rule namely that
the total proton spin must come out to be $\frac12$, whatever the various
contributions are. 
In the chiral bag model, where the mesonic
degrees of freedom also play an important role, the proton spin is
described by the following contributions
\begin{equation}
\textstyle \frac12 = \frac12 \Sigma +  L_Q + \Gamma + L_G + L_M,
\end{equation}
where $L_M$, the orbital angular momentum  of the
mesons, has to be added to Eq.(\ref{QG}). The proton spin is generated by
quantizing the collective rotation associated with the zero modes
of the classical soliton solution of the model Lagrangian. To the
collective rotation, each constituent responds with the
corresponding moment of inertia. The moments of inertia of the
quarks and mesons, ${\cal I}_Q$ and ${\cal I}_M$, have been
extensively studied in the literature\cite{PR88}. Substitution of
the color electric and magnetic fields, given by Eqs.(\ref{vecE})
and (\ref{vecB}) respectively, into Eq.(\ref{Jg}) defines a new
moment of inertia of the static gluon fields with respect to the
collective rotation as
\begin{equation}
\langle {\bf J}_G \rangle = - {\cal I}_G {\bf \omega},
\label{IG}
\end{equation}
where the expectation value is taken keeping only the exchange terms,
and ${\bf\omega}$ is the classical angular velocity of the
collective rotation.

We show in Fig.1(a) and (b) the gluon moment of inertia evaluated
by using the color electric fields with $Q_R(r)$ and $Q_0(r)$. In
the case of $Q_R(r)$, ${\cal I}_G$ is positive for all bag radii
and comparable in size to ${\cal I}_Q$, the quark moment of
inertia. On the other hand, $Q_0(r)$ results in a negative ${\cal
I}_G$. This ``negative" moment of inertia may appear to be bizarre
but it may not be a problem from the conceptual point of view. The
${\cal I}_G$ defined by Eq.(\ref{IG}) can be interpreted as the
one-gluon exchange correction to the corresponding quantity of the
quark phase, which is still positive anyway. The point is that the
spin fractionizes in the same way as the moment of inertia does.
This means that we have
\begin{eqnarray}
L_Q+\textstyle\frac12 \Sigma &=&
\frac{{\cal I}_Q}{2({\cal I}_Q+{\cal I}_G+{\cal I}_M)},
\nonumber\\
L_G+\Gamma &=&
\frac{{\cal I}_G}{2({\cal I}_Q+{\cal I}_G+{\cal I}_M)},
\\
L_M &=&
\frac{{\cal I}_M}{2({\cal I}_Q+{\cal I}_G+{\cal I}_M)}.
\nonumber
\end{eqnarray}
Each fraction as a function of the bag radius is presented in 
the small boxes inside each figure. Note in the case of adopting 
$Q_R(r)$ that at the
large bag limit the proton spin is equally carried by quarks and
gluons somewhat like the momentum of the proton. The negative
${\cal I}_G$ obtained with $Q_0(r)$, thus, yields a scenario
where the gluons are anti-aligned with the proton spin.

The dashed and dash-dotted curves in Figs.2(a) and (b) show
the values for $\Gamma_\lambda$ and $\Gamma^\prime_\lambda$.
For comparison, we draw $\frac12\Sigma$ by a solid curve.
Note that, because of the difference in ${\cal I}_G$, even
$\frac12\Sigma$ is different according to which $Q_\lambda$ is
used. Again, both $\Gamma_0$ and $\Gamma^\prime_0$ are
anti-aligned with the proton spin. Note of course that the
negative $\Gamma_0^\prime$ is apparently at variance with the
general belief that the anomaly is to cure the proton spin
problem.

To conclude, we show in Figures 3(a) and (b) the flavor-singlet
axial current including the $U_A(1)$ anomaly given by
\begin{equation}
a_0 = \Sigma - (N_f\alpha/2\pi) \Gamma^\prime_\lambda.
\end{equation}
For simplicity, we neglect other contributions to $a_0$ studied
in \cite{LMPRV99}. They show that the positive $\Gamma$ is
consistent with the small $a_0$ measured in the EMC experiments.
The radius dependence of each component may be viewed as gauge
dependence both in color gauge symmetry and in the ``Cheshire Cat"
gauge symmetry discussed by Damgaard, Nielsen and
Sollacher~\cite{CC}.

\section*{Acknowledgments}
The work of HJL, DPM and BYP is supported in part by the KOSEF grant
1999-2-111-005-5. The work of VV is partially supported by DGICYT grant PB97-1127.
We would like to thank KIAS where this work was initiated for
the generous support.

\newpage

\begin{figure}
\caption{The moment of inertia associated with the collective rotation
as a function of the bag radius and the proton spin fraction carried by each
constituents. In the calculation, we have used (a) the ``confined" color
electric field with $Q_R(r)$ and (b) the conventional one with $Q_0(r)$.}
\centerline{\epsfig{file=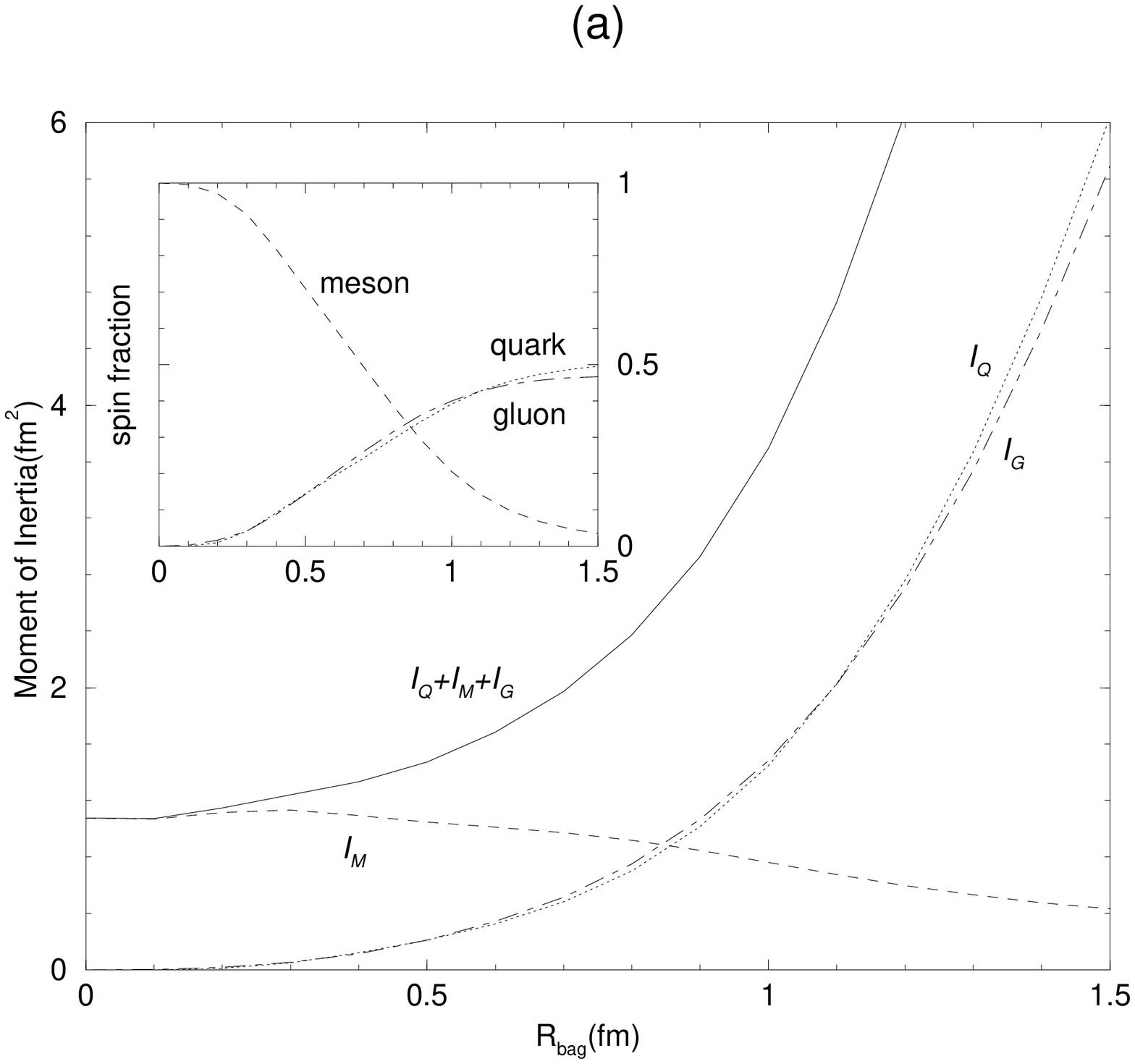, width=7.5cm, angle=270}
\epsfig{file=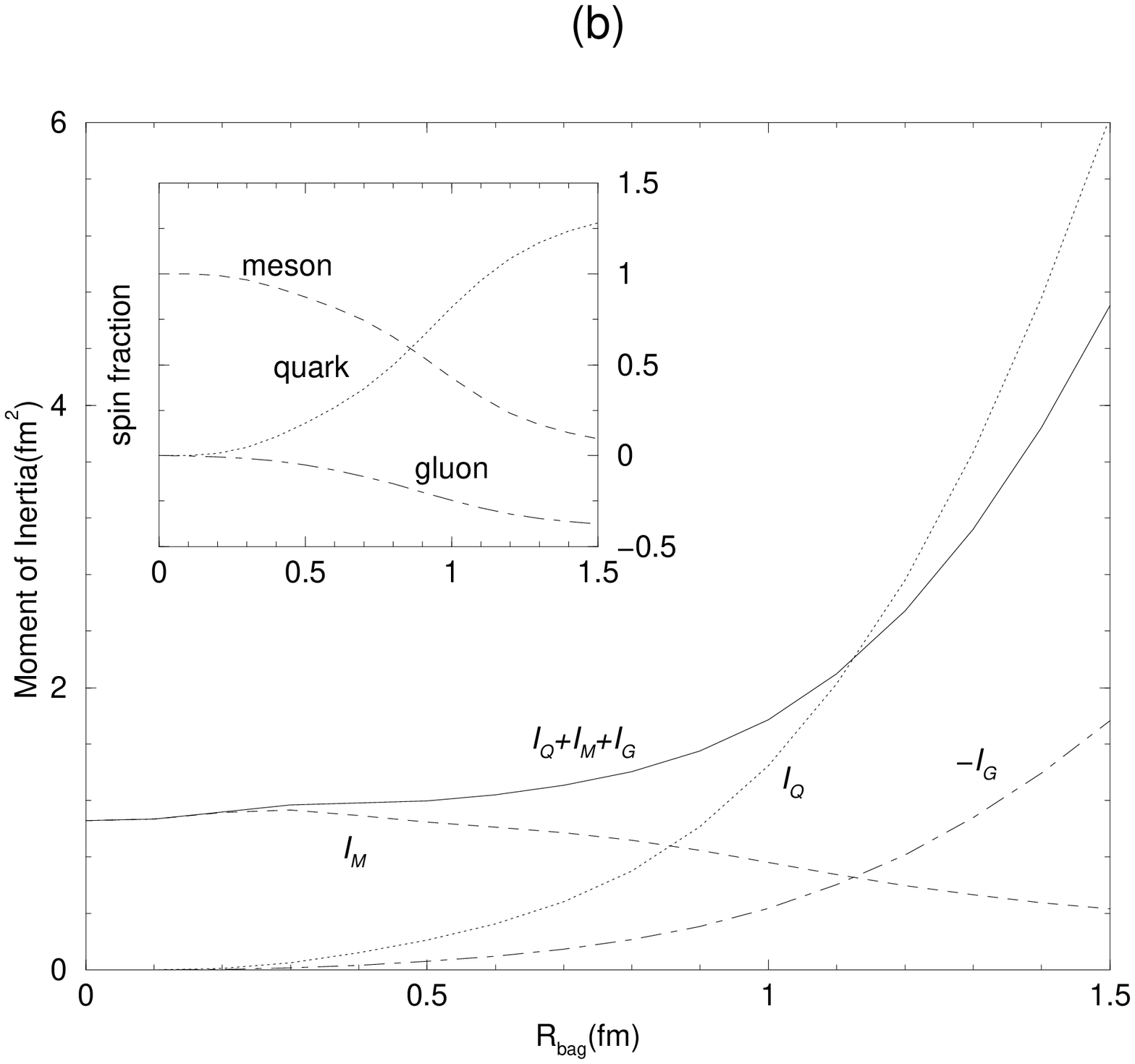, width=7.5cm, angle=270}}
\end{figure}

\begin{figure}
\caption{The gluon spin $\Gamma$ as a function of the bag radius.
(a) and (b) are obtained with the color electric fields
explained in Fig. 1.}
\centerline{\epsfig{file=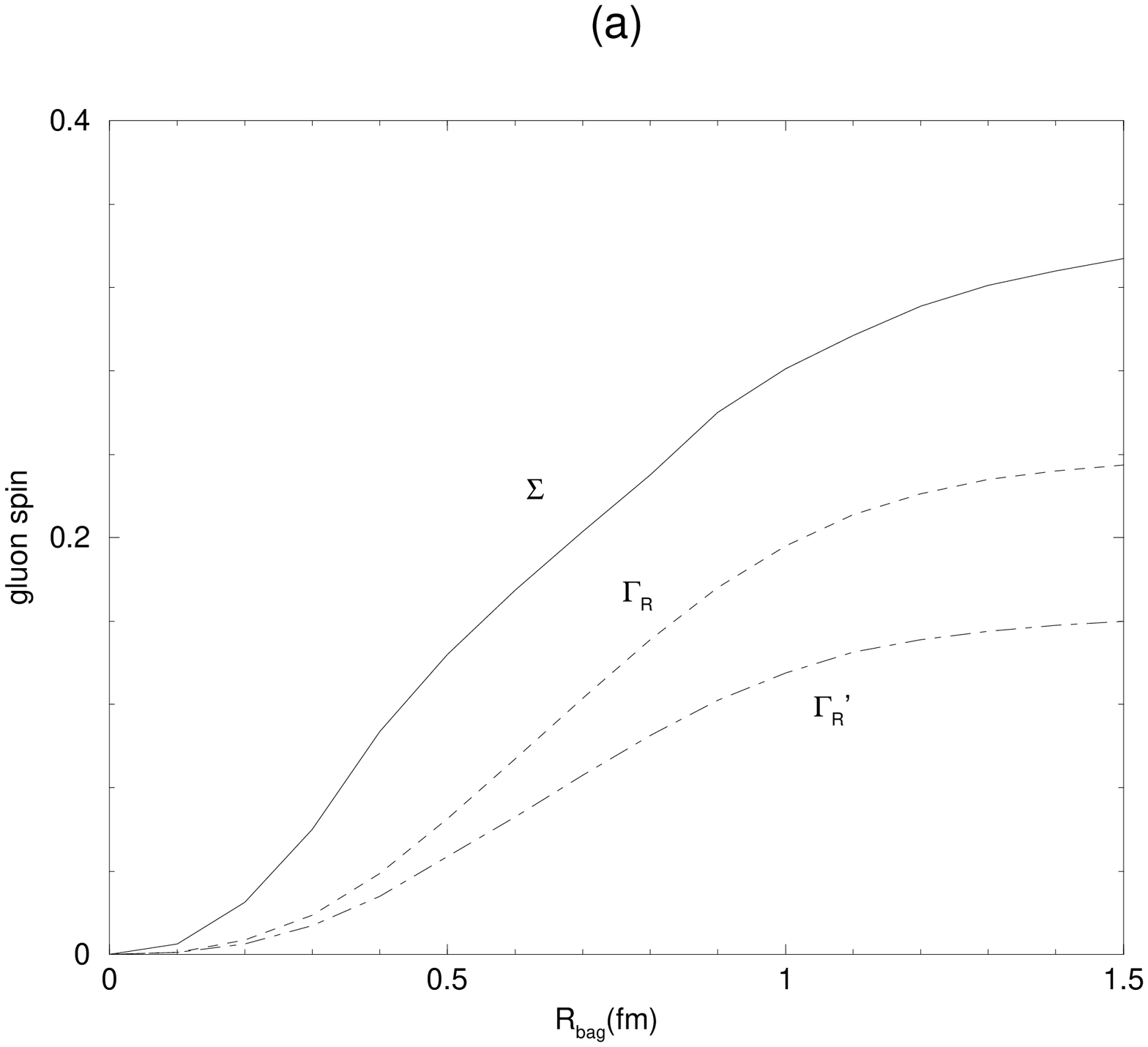, width=7.5cm, angle=270}
\epsfig{file=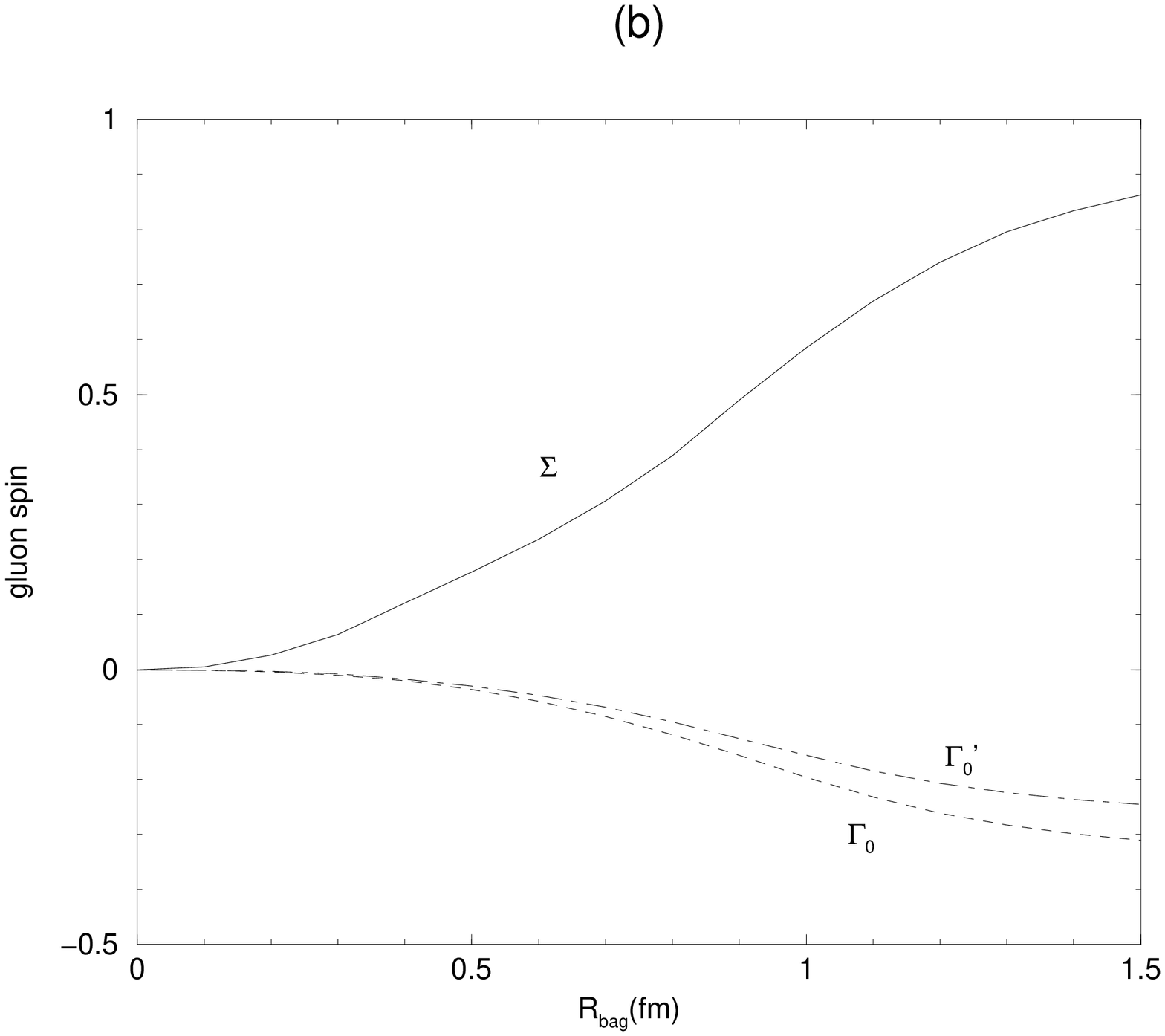, width=7.5cm, angle=270}}
\end{figure}

\begin{figure}
\caption{The flavor singlet axial current $a_0$
as a function of the bag radius.
(a) and (b) are obtained with the color electric fields
explained in Fig. 1.}
\centerline{\epsfig{file=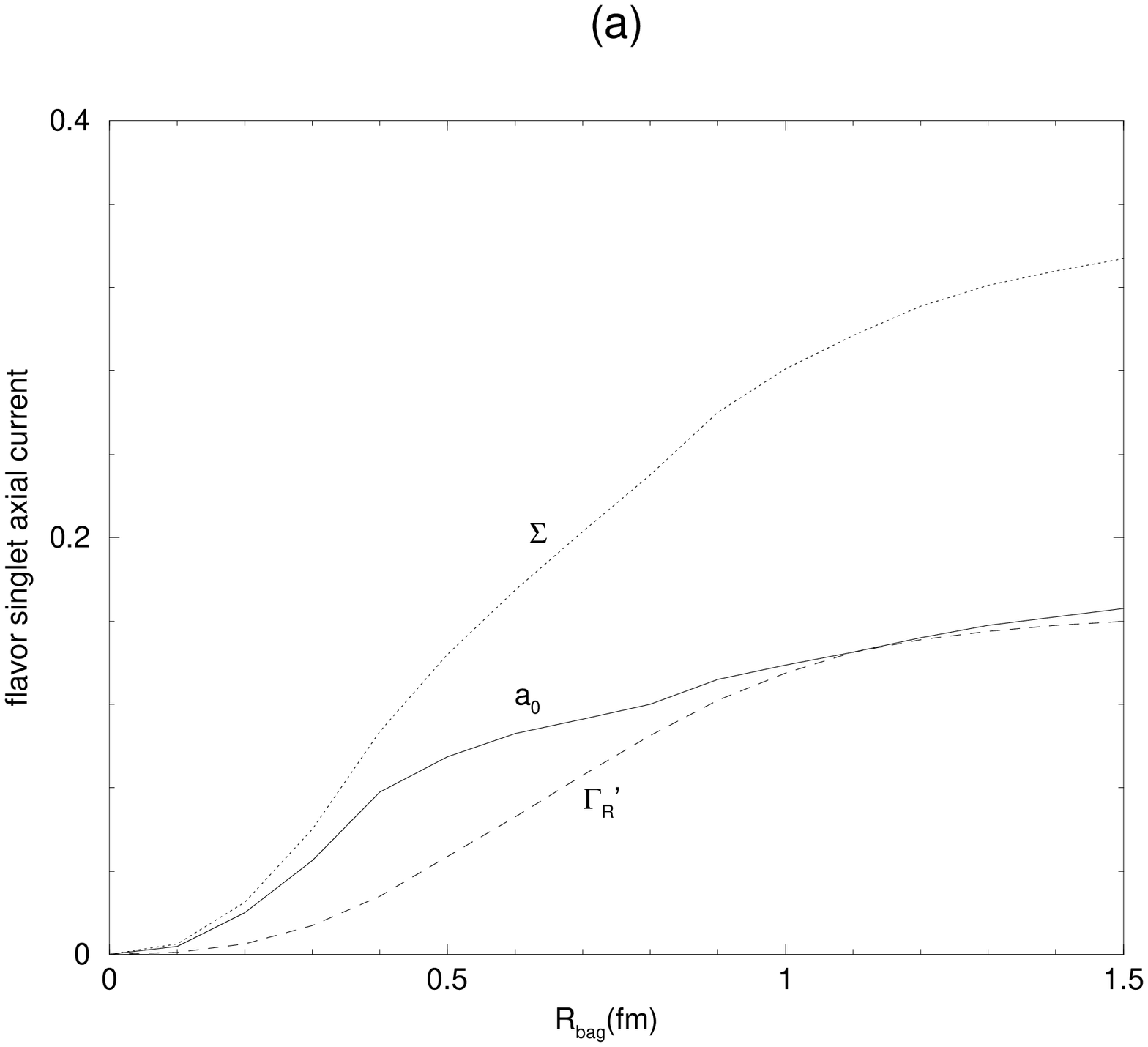, width=7.5cm, angle=270}
\epsfig{file=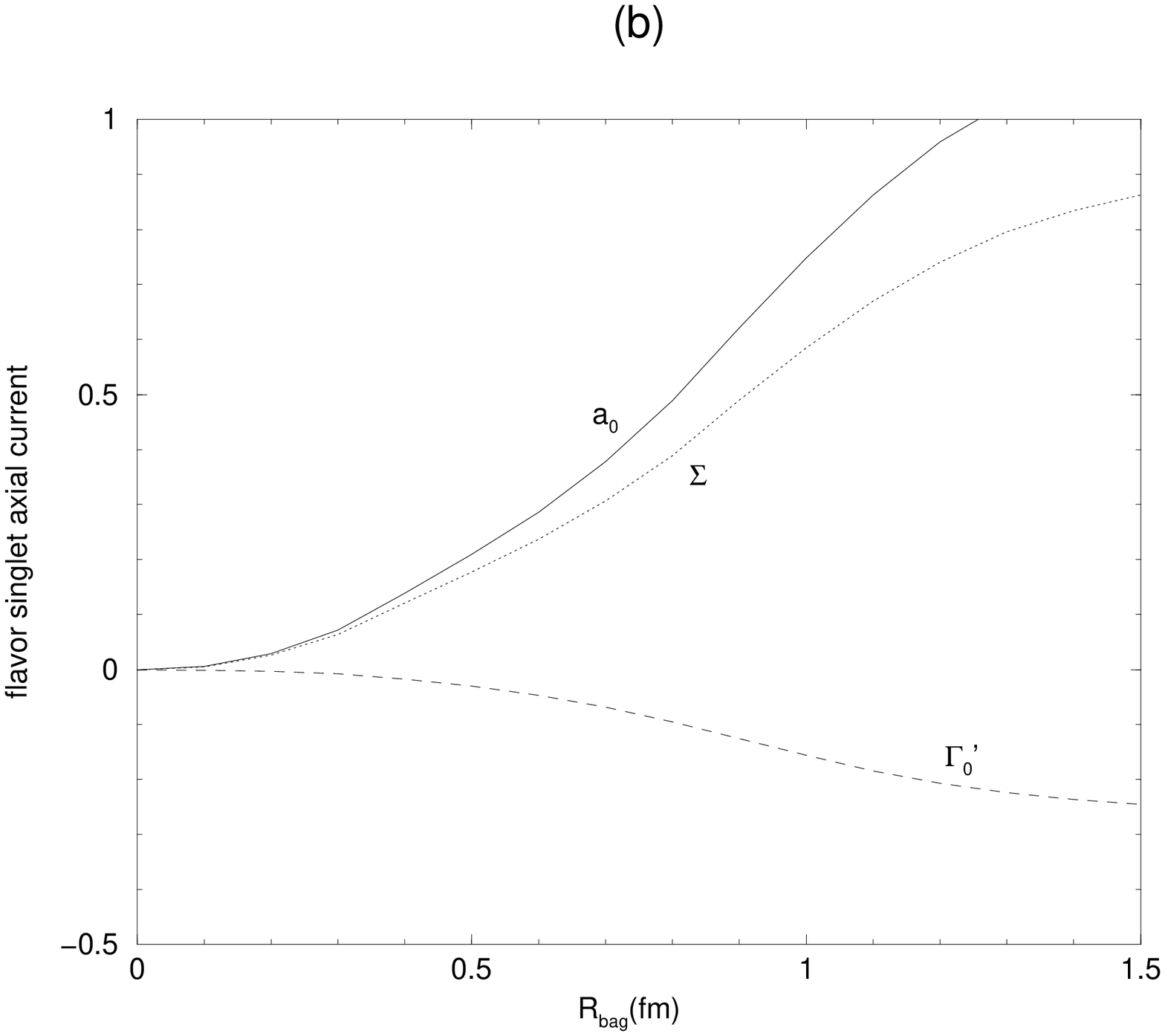, width=7.5cm, angle=270}}
\end{figure}

\end{document}